\documentclass[aps,prl, reprint, 
showpacs,showkeys,superscriptaddress,preprintnumbers,nofootinbib]{revtex4-1} 

\usepackage{amsmath,amssymb,bm,mathrsfs}
\usepackage{hyperref}
\usepackage{url}
\usepackage{graphicx}

\begin{document}

\preprint{\tt UT-18-13, IPMU 18-0111}

\title{Limit on the Axion Decay Constant from the Cooling Neutron Star in Cassiopeia A} 

\author{Koichi Hamaguchi}
\email{hama@hep-th.phys.s.u-tokyo.ac.jp}
\affiliation{Department of Physics, University of Tokyo, 
Tokyo 113--0033,
Japan}
\affiliation{
Kavli Institute for the Physics and Mathematics of the Universe (Kavli
IPMU), University of Tokyo, Kashiwa 277--8583, Japan
}

\author{Natsumi Nagata}
\email{natsumi@hep-th.phys.s.u-tokyo.ac.jp}
\affiliation{Department of Physics, University of Tokyo, 
Tokyo 113--0033,
Japan}

\author{Keisuke Yanagi}
\email{yanagi@hep-th.phys.s.u-tokyo.ac.jp}
\affiliation{Department of Physics, University of Tokyo, 
Tokyo 113--0033,
Japan}

\author{Jiaming Zheng}
\email{zhengjm3@hep-th.phys.s.u-tokyo.ac.jp}
\affiliation{Department of Physics, University of Tokyo, 
Tokyo 113--0033,
Japan}

\begin{abstract}
 The observed rapid cooling of the neutron star (NS) located at the center of the supernova remnant Cassiopeia A (Cas A) can be explained in the minimal NS cooling scenario. This consequence may be changed if there exists an extra cooling source, such as axion emission. In this work, we study the Cas A NS cooling in the presence of axion emission, taking account of the temperature evolution in the whole life of the Cas A NS. We obtain a lower limit on the axion decay constant, $f_a\gtrsim (5-7)\times 10^8$ GeV, if the star has an envelope with a thin carbon layer.
This is as strong as existing limits imposed by other astrophysical observations such as SN1987A.

\end{abstract}

\maketitle

\section{Introduction}

The Cassiopeia A (Cas A) is a supernova remnant in the Cassiopeia
constellation. This supernova may be identical to the \textit{3 Cassiopeiae}
recorded by John Flamsteed on August 16, 1680
\cite{1980JHA....11....1A, *1980Obs...100....3K, *1980Natur.285..132H},
which is consistent with the supernova explosion date estimated from the
remnant expansion: $1681\pm 19$ \cite{Fesen:2006zma}. In 1999, the
\textit{Chandra} X-ray observatory discovered a hot point-like source in
the center of the supernova remnant \cite{1999IAUC.7246....1T}, which is
now identified as a neutron star (NS). 

Given the distance to Cas A, $d = 3.4^{+0.3}_{-0.1}$~kpc
\cite{Reed:1995zz}, the NS radius can be determined by measuring
the X-ray spectrum thermally emitted from the NS. With the black-body
and hydrogen atmosphere models, a rather small radius of the X-ray emission
area was obtained---about 0.5 and 2 km, respectively \cite{Pavlov:1999tca, *Chakrabarty:2000ps, *Pavlov:2003eg}. This implied a
hot spot on the NS surface. On the other hand, a lack of the observation of
pulsations in the X-ray flux \cite{Murray:2001fy, *Mereghetti:2001cf}
indicates that the X-ray emission comes from the whole surface, which is
incompatible with the above observation. This contradiction was resolved 
by Heinke and Ho, who found that a carbon atmosphere model with low
magnetic field gave a good fit to the X-ray spectrum with a
typical size of the NS radius ($\gtrsim 10$~km)
\cite{Ho:2009mm}. Moreover, they observed that the surface temperature
of the NS evaluated with the carbon atmosphere model was decreasing over
the years, which provided the first direct observation of NS cooling
\cite{Heinke:2010cr}. 

In the standard NS cooling scenario (see, \textit{e.g.},
Refs.~\cite{Yakovlev:2004iq, *Page:2005fq}), a
young NS like the Cas A NS cools down mainly through neutrino
emission. Soon after the work by Heinke and Ho \cite{Heinke:2010cr},
the authors in Refs.~\cite{Page:2010aw, Shternin:2010qi} tried
to fit the observed data with the standard NS cooling model.\footnote{See Refs.~\cite{Yang:2011yg, *Negreiros:2011ak, *Blaschke:2011gc, *Noda:2011ag, *Sedrakian:2013xgk, *Blaschke:2013vma, *Bonanno:2013oua, *Leinson:2014cja, *Taranto:2015ubs, *Noda:2015pvn, *Grigorian:2016leu} for other possible mechanisms.} They argued
that the cooling rate of the Cas A NS was so rapid that the ``slow'' neutrino
processes could not fit the cooling curve, while the ``fast'' neutrino
processes predicted a temperature smaller than the observed one
around the time $t \simeq 320$~years. It was then concluded that the
neutron triplet superfluid transition should occur around this time so that
neutrino emission was accelerated through the breaking of neutron
Cooper pairs and their subsequent reformation, dubbed as the
``pair-breaking and formation (PBF)'' process \cite{Flowers:1976ux,
*Voskresensky:1987hm}. In addition, proton superconductivity should
operate well before $t \simeq 320$~years in order to suppress neutrino
emission in the early times, which results in a steep decrease in
temperature right after the onset of the PBF process. The observed
cooling curve of the Cas A NS was found to be fitted quite well under
these conditions.

This conclusion may be altered if there is an additional source of NS
cooling. One of the most well-motivated candidates for such a cooling
source is the emission of axions \cite{Weinberg:1977ma, *Wilczek:1977pj},
the Nambu-Goldstone bosons associated with the Peccei-Quinn symmetry
\cite{Peccei:1977hh, *Peccei:1977ur}. 
The axions are emitted through the axion PBF and bremsstrahlung processes caused by the couplings to nucleons.
Indeed, there are several studies that
discuss their effects on the Cas A NS cooling. 
A detailed study on the axion emission processes and their consequences on
NS cooling was performed in Ref.~\cite{Sedrakian:2015krq}, where
predicted cooling curves were compared with the temperature data of
young NSs including the Cas A NS. However, only the average
temperature of the Cas A NS was concerned and no attempt was made to fit
the slope of its cooling curve. In Ref.~\cite{Leinson:2014ioa},
the axion-neutron PBF process was utilized to enhance the cooling rate so that
the slope of the Cas A NS cooling curve was reproduced. This
analysis focused on the time around which the neutron
superfluid transition was supposed to occur, with the
axion-proton PBF and axion bremsstrahlung processes neglected; 
especially, there was no discussion on the
temperature evolution at $t\lesssim 300$~years. 

In this work, we study the Cas A NS cooling in the presence of axion emission, taking account of the temperature evolution
in the whole lifetime of the Cas A NS. It is found that for a
sufficiently small axion-nucleon coupling, the observed cooling curve
can still be fitted if we take a moderate gap for neutron triplet pairing
and a large gap for proton singlet pairing. This success is, however,
spoiled once the axion coupling exceeds a certain value, and therefore imposes
a lower limit on the axion decay constant. This new limit turns out to be
as strong as the existing limits given by other astrophysical observations such as SN1987A.

\section{Standard NS Cooling and Cas A NS}

The Cas A NS cooling data collected so far \cite{Ho:2014pta} clearly
shows that the temperature is decreasing at a constant rate. 
The measured X-ray spectrum is fitted with an atmosphere model, and the radius $R$ and mass $M$ of the Cas A NS are inferred through their effects on the brightness and gravitational redshift.
In Ref.~\cite{Ho:2014pta}, the authors performed such a fit using
a non-magnetic carbon atmosphere model
\cite{Ho:2009mm} and obtained $M \simeq (1.4 \pm 0.3) M_{\odot}$.

In the standard NS cooling scenario \cite{Yakovlev:2004iq, 
*Page:2005fq}, the NS cooling proceeds via the emission of
neutrinos and photons. The former dominates over the latter in the earlier
epoch ($t \lesssim 10^{5}$~years). Various processes participate in
neutrino emission, such as the direct Urca process, the modified Urca
process, bremsstrahlung, and the PBF process. The direct Urca process,
comprised of the $\beta$-decay and inverse
decay processes, is called the ``fast'' process.
If this occurs, a NS cools quite rapidly. However, this process can
occur only at very high density regions \cite{Lattimer:1991ib}, which
can be achieved only for a heavy NS. For instance, the
Akmal-Pandharipande-Ravenhall (APR) equation of state
\cite{Akmal:1998cf} allows the direct Urca process only for $M \gtrsim
1.97M_{\odot}$, which is well above the Cas A NS mass estimated in
Ref.~\cite{Ho:2014pta}. Thus, we can safely assume that the fast
process never occurs in the Cas A NS, as in the minimal
cooling paradigm \cite{Page:2004fy, Page:2009fu}. In this case, the
neutrino emission proceeds through the ``slow'' processes such as the
modified Urca and bremsstrahlung processes. 
In the absence of nucleon pairings, the
neutrino luminosity $L_\nu$ caused by these processes is expressed as
$L_\nu \simeq L_9 T_9^8$ with the internal temperature $T_9 \equiv
T/(10^9~\text{K})$ and the coefficient $L_9 \sim
10^{40}~\text{erg}\cdot\text{s}^{-1}$ \cite{Yakovlev:2004iq,
*Page:2005fq}. 

In general, a NS is brought into an isothermal state with
relaxation time $\sim 10$--100~years \cite{1994ApJ...425..802L,
*Gnedin:2000me}, and in fact the Cas A NS is very likely to be thermally
relaxed \cite{Yakovlev:2010ed}. For an isothermal NS, the evolution of
the internal temperature $T$ is determined by the thermal balance equation 
\begin{equation}
 C \frac{dT}{dt} = -L_\nu -L_{\text{cool}} ~,
\label{eq:balanceeq}
\end{equation}
where $C$ is the total stellar heat capacity and $L_{\text{cool}}$
denotes the luminosity caused by potential extra cooling sources. We have dropped
the photon luminosity since this is much smaller than $L_\nu$ for a
young NS like the Cas A NS. The heat capacity $C$
has temperature dependence of the form $C = C_9 T_9$ with $C_9 \sim
10^{39}~\text{erg} \cdot \text{K}^{-1}$ \cite{Yakovlev:2004iq,
*Page:2005fq}. 
If $L_{\text{cool}} = 0$, Eq.~\eqref{eq:balanceeq} leads to 
\begin{equation}
 T_9 = \biggl(\frac{C_9 \cdot 10^9~\text{K}}{6L_9
  t}\biggr)^{\frac{1}{6}} 
\sim  \biggl(\frac{1~\text{year}}{ t}\biggr)^{\frac{1}{6}}
~,
\label{eq:t9}
\end{equation}
where we have assumed that the initial temperature is much larger than
that at the time of interest. 

The NS surface is insulated from the hot interior by its envelope. 
For a non-magnetic iron envelope at temperatures as high as the Cas A NS
temperature, the relation between the surface temperature $T_s$ and the
internal temperature $T$ is approximated by \cite{1983ApJ...272..286G}
\begin{equation}
 T_9 \simeq 0.1288 \times \biggl(\frac{T_{s6}^4}{g_{14}}\biggr)^{0.455} ~,
\label{eq:GPE}
\end{equation}
with $g_{14}$ the surface gravity in units of $10^{14}~\text{cm}\cdot
\text{s}^{-2}$ and $T_{s6} \equiv T_s/(10^6~\text{K})$. 
A more accurate relation, which is also applicable to the case where a
sizable amount of light elements exist in the envelope
(characterized by the parameter $\eta \equiv g_{14}^2
\Delta M/M$ with $\Delta M$ the mass of the light elements),
can also be found in the literature \cite{Potekhin:1997mn}. 

The cooling rate of the Cas A NS observed in Ref.~\cite{Heinke:2010cr,Ho:2014pta}
was about 3--4\% in ten years around $t \simeq 320$~years.\footnote{See also \cite{Elshamouty:2013nfa,*Posselt:2013xva} for possible uncertainties.} On the other hand, 
from Eq.~\eqref{eq:t9} and Eq.~\eqref{eq:GPE}, we find that the surface
temperature goes as $T_s \propto t^{0.09}$,
which results in only
$0.3$\% decrease in temperature in ten years. Hence, the slow
neutrino emission cannot explain the observed rapid cooling of the Cas A NS. 
 
This difficulty can be resolved with the help of superfluidity in the
NS (for a review, see Ref.~\cite{Page:2013hxa}). It is known that the onset of Cooper pairing of nucleons triggers the rapid PBF emission of neutrino while suppresses other emission processes which these nucleons participate in~\cite{Flowers:1976ux,
*Voskresensky:1987hm}. The PBF lasts only for a short
time---to explain the rapid cooling of the Cas A NS by this PBF process, therefore, the phase transition of the neutron triplet
pairing should occur just before $t\simeq 320$~years. This condition implies
that the critical temperature of this phase transition, $T_c^{(n)}$,
should agree to the internal temperature around this time; thus,
$T_c^{(n)}$ is fixed via Eq.~\eqref{eq:t9}. One also finds that the
resultant cooling rate increases as $T_c^{(n)}$ gets larger, which is
achieved with a smaller $L_9$ according to
Eq.~\eqref{eq:t9}. The reduction in $L_\nu$ can be realized again with the
aid of Cooper pairing---with proton pairings formed, the neutrino emission
processes which contain protons are suppressed by the proton gap, which
then results in a small $L_9$. A small $L_9$ also ensures that the NS was not overcooled by the time of observation.

Indeed, the authors in Refs.~\cite{Page:2010aw, Shternin:2010qi} found
that the rapid cooling of the Cas A NS can be explained in
the minimal cooling scenario with an appropriate choice of
$T_c^{(n)}$ and a sufficiently large proton pairing gap. For instance,
it is shown in Ref.~\cite{Page:2010aw} that the observed data points are
fitted quite well 
for $T_c^{(n)} \simeq 5.5 \times 10^8$~K and $\Delta M = 5 \times
10^{-13} M_{\odot}$, where the CCDK model for proton gap
\cite{Chen:1993bam} and the APR equation of state \cite{Akmal:1998cf}
are adopted.  
In Refs.~\cite{Page:2012se,Shternin:2010qi}, it was shown 
that the temperature observations of other NSs can also be
fitted by cooling curves with pairing models required by the rapid cooling of the Cas A NS. Because of its simplicity, the minimal cooling scenario is a very promising candidate of the correct NS cooling model.
In the rest of this paper, we will consider the compatibility between the minimal NS
cooling model and axion models by looking for the highest axion decay constant
$f_a$ with which the rapid Cas A NS cooling cannot be fitted by any pairing model. This
serves as a lower bound of $f_a$ under the assumption that the
minimal NS cooling model describes the cooling of NS correctly.

\section{Axion Emission from Neutron Stars}

The discussion in the last section would be changed if there is an
additional cooling source, \textit{i.e.}, if $L_{\text{cool}} \neq 0$ in
Eq.~\eqref{eq:balanceeq}. In this case, the temperature at $t\simeq
320$~years is predicted to be lower than that in the minimal cooling
scenario.
However, 
the observed surface temperature of the Cas A NS, $T_s \simeq 2 \times 10^6$~K, implies
$T \simeq 4\times 10^8$~K
(see Eq.~\eqref{eq:GPE}), and $T_c^{(n)}$ needs to be larger than this
value in order for the PBF process to operate at $t\simeq 320$ years. In other words,
if we fix $T_c^{(n)} \simeq 5.5 \times 10^8$~K in the case of
$L_{\text{cool}} \neq 0$, the rapid cooling due to the PBF process
has occurred much before $t \simeq 320$~years---then, the rapid
cooling would have already ceased and/or the present surface temperature
would be much lower than the observed value. Accordingly, we may obtain
a constraint on extra cooling sources from the Cas A NS cooling data. 

To discuss this possibility, in this work, we take axion
\cite{Weinberg:1977ma, *Wilczek:1977pj} as a concrete example for a
cooling source. Axions are emitted out of NSs through their couplings to
nucleons. The axion-nucleon couplings have the form
\begin{equation}
 {\cal L}_{\text{int}} = \sum_{N =p,n} \frac{C_N}{2 f_a} \bar{N} \gamma^\mu
  \gamma_5 N \partial_\mu a ~,
\label{eq:axion_nucleon}
\end{equation}
where $f_a$ is the axion decay constant. The coefficients $C_N$ are
expressed in terms of the axion-quark couplings $C_q$ (having the same
form as in Eq.~\eqref{eq:axion_nucleon}), the quark masses $m_q$, and
the spin fractions $\Delta q^{(N)}$ defined by $2 s_\mu^{(N)} \Delta
q^{(N)} \equiv \langle N| \bar{q} \gamma_\mu \gamma_5 q |N\rangle$ with
$s_\mu^{(N)}$ the spin of the nucleon $N$. At the leading order in the
strong coupling constant $\alpha_s$, we have $C_N = \sum_{q} (C_q
-m_*/m_q) \Delta q^{(N)}$ with $m_* \equiv m_um_dm_s/(m_u m_d + m_d m_s + m_u
m_s) $. QCD corrections to this formula are discussed in
Ref.~\cite{diCortona:2015ldu}; for instance, in the case of the KSVZ
axion ($C_q = 0$) \cite{Kim:1979if, *Shifman:1979if}, we have $C_p =
-0.47(3)$ and $C_n =-0.02(3)$, while for the DFSZ axion
\cite{Zhitnitsky:1980tq, *Dine:1981rt} ($C_{u,c,t} = \cos^2 \beta/3$ and
$C_{d,s,b} = \sin^2 \beta/3$ with $\tan\beta$ the ratio of the vacuum
expectation values of the two doublet Higgs fields, $\tan \beta \equiv
\langle H_u\rangle /\langle H_d\rangle$),
 $C_p = -0.182(25) - 0.435 \sin^2
\beta$ and $C_n = -0.160(25) + 0.414 \sin^2 \beta$
\cite{Patrignani:2016xqp}, where $\Delta u^{(p)} = \Delta d^{(n)} =
0.897(27)$, $\Delta d^{(p)} = \Delta u^{(n)} = -0.376(27)$, and $\Delta
s^{(p)} = \Delta s^{(n)} = -0.026(4)$ are used.

The axion-nucleon couplings  induce axion
emission via the PBF and bremsstrahlung processes, which have been
studied so far in the literature \cite{Iwamoto:1984ir, *Nakagawa:1987pga,
*Nakagawa:1988rhp, *Iwamoto:1992jp, *Umeda:1997da, Leinson:2014ioa,
Sedrakian:2015krq, Paul:2018msp}. We have modified the public code
\texttt{NSCool} \cite{NSCool} to implement these processes and use it to
compute the luminosity of axion and its effect on the  NS
cooling curves. We adopt the APR equation of state \cite{Akmal:1998cf}
and fix the NS mass to be $M = 1.4 M_\odot$ in this work.

\begin{figure}
{\includegraphics[clip, width = 0.45 \textwidth]{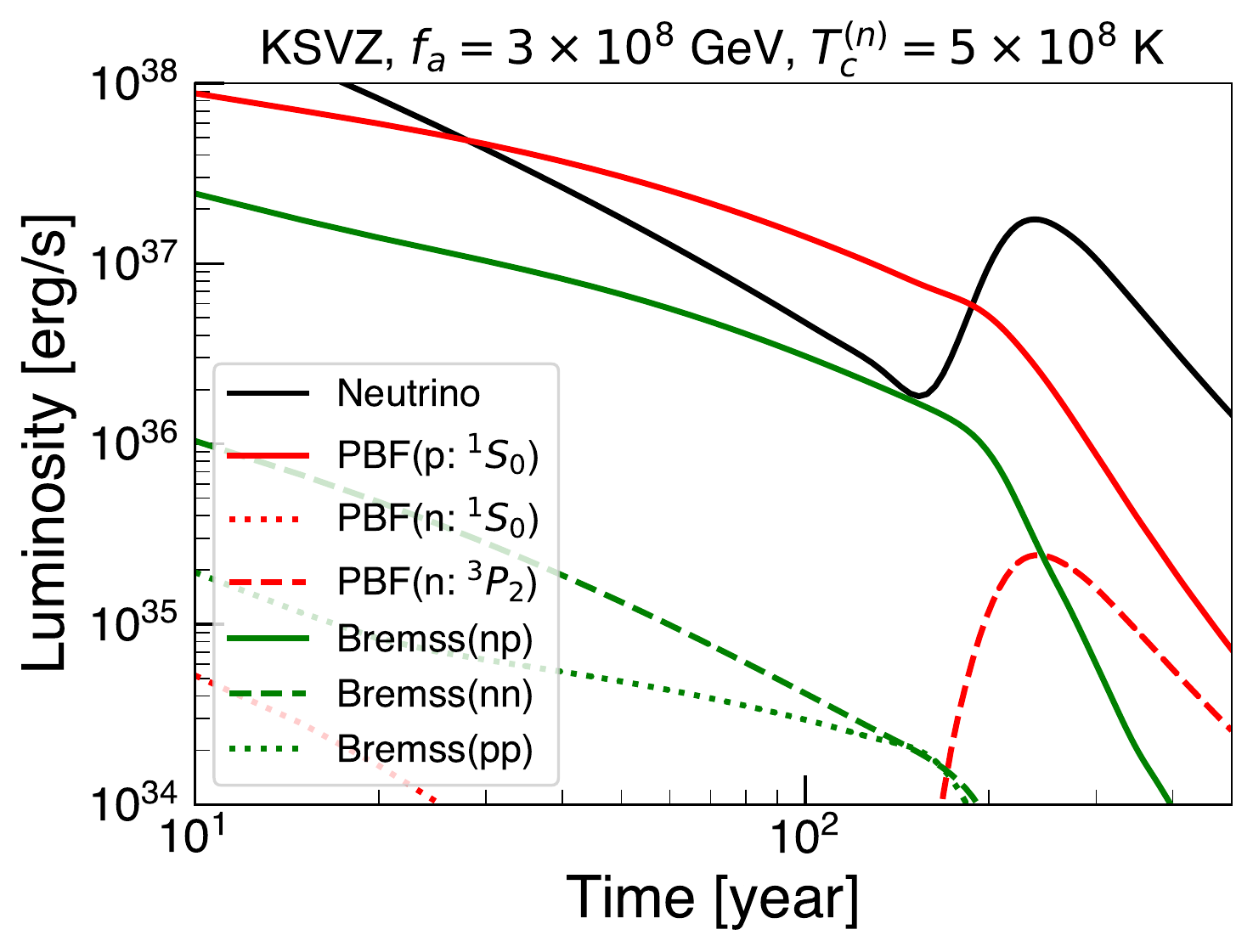}} 
\caption{
Luminosity of each axion emission  (red and green) and the total 
neutrino emission (black) processes as a function of time.  
}
\label{fig:Lumiksvz}
\end{figure}

In Fig.~\ref{fig:Lumiksvz}, we show luminosities of various axion
emission processes in the KSVZ model with $f_a=3\times 10^8$~GeV as functions of time (red and green).
For comparison, we also show the total luminosity of neutrino emission (black). 
We use the SFB model~\cite{Schwenk:2002fq} for the gap of singlet neutron pairings.
Our analysis is insensitive to this choice.
For the proton singlet pairings, the CCDK model~\cite{Chen:1993bam} is chosen because it has the largest gap in the NS core among
those presented in Ref.~\cite{Page:2004fy}---this results in a strong
suppression of neutrino emission before the onset of the neutron triplet Cooper pairing
and thus improves the fit onto the
observed Cas A NS data \cite{Page:2010aw} as we discussed
above. 
Note that a large gap for the proton singlet pairing also suppresses the axion emission, and therefore the CCDK model gives a conservative bound.
On the contrary, there are large uncertainties in choosing a model of neutron
triplet pairings. We thus model this gap with a Gaussian shape with
respect to the neutron Fermi momentum  and regard its height, width, and
central position  as free parameters. In Fig.~\ref{fig:Lumiksvz}, we take
$T_c^{(n)} = 5 \times 10^8$~K. The instantaneous increase in
luminosity at $t \simeq 300$ years for the neutrino emission, as well as
the axion emission in the PBF process, is due to the
formation of neutron triplet pairings. As we see from this plot, the
axion emission via the proton PBF and proton-neutron bremsstrahlung
processes is as strong as the neutrino emission before the formation of
neutron triplet parings. In particular, the emission
via the proton PBF dominates over other axion emission processes in this case because
it is suppressed by less powers of $T_{\text{core}}$ resulting from a smaller number of
 states involved in the process. This allows us to set stringent bounds even
on the KSVZ model where $|C_n|$ is vanishingly small. If $|C_n|$ is sizable as in the DFSZ model, the
neutron bremsstrahlung process is also significant.

\section{Limit on axion decay constant}

\begin{figure}
{\includegraphics[clip, width = 0.45 \textwidth]{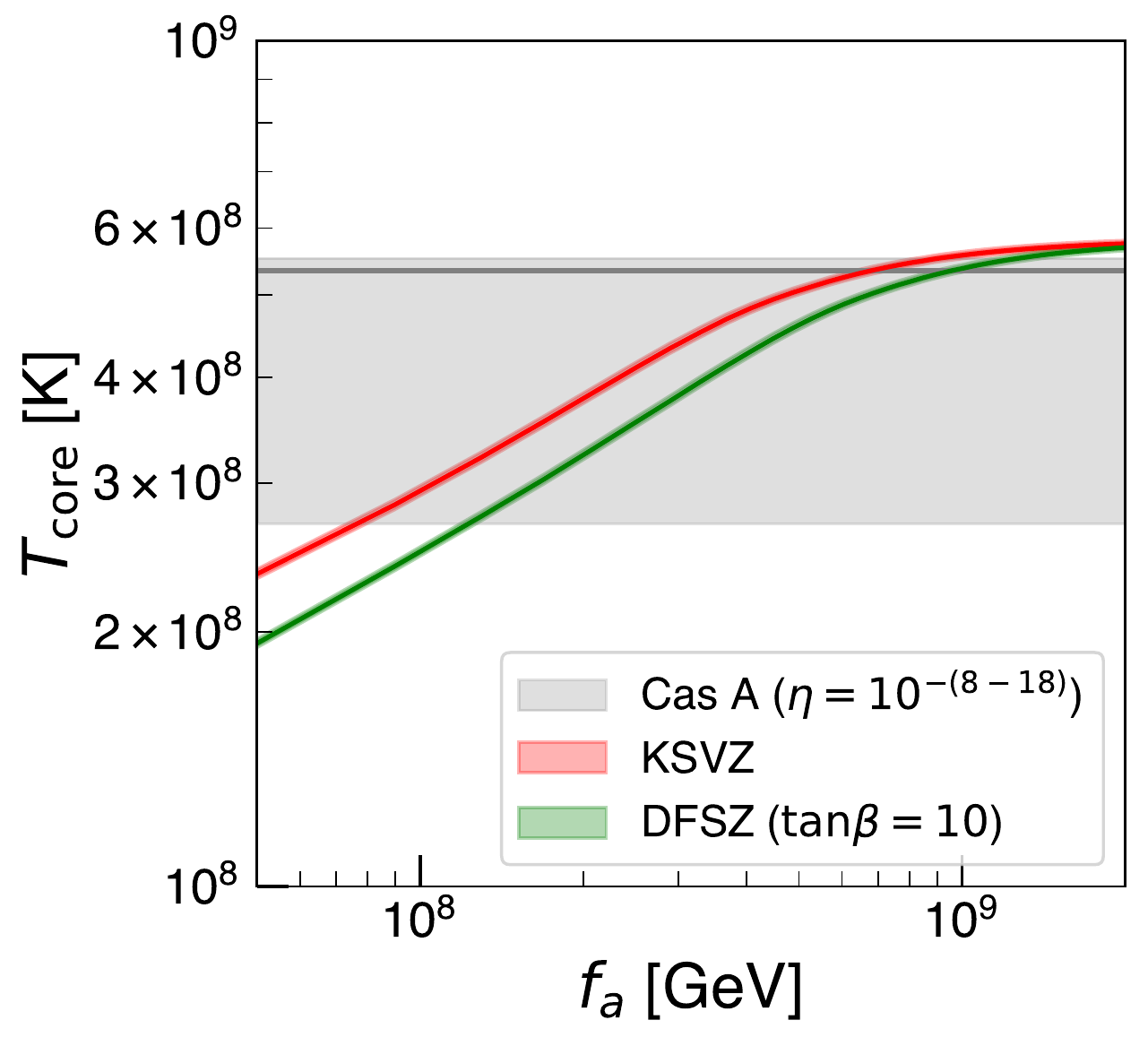}} 
\caption{
The core temperature $T_{\mathrm{core}}$ at $t = 300$--338~years against $f_a$, without neutron triplet superfluidity. The gray shaded region reflects the uncertainty of $T_{\mathrm{core}}$ due to $\eta$ and the grey solid line corresponds to $\eta = 5\times 10^{-13}$.
}
\label{fig:tvsfa}
\end{figure}

Now let us study the effect of the axion emission processes on the NS temperature evolution. In Fig.~\ref{fig:tvsfa}, we show the core temperature
$T_{\mathrm{core}}$ at the time of the Cas A NS age on January 30, 2000 ($t
= 300$--338 years) as functions of $f_a$ for the KSVZ and DFSZ
($\tan\beta = 10$) models in the red and green bands, respectively, with
the bands reflecting the uncertainty in the NS
age. Since we are interested in the drop in the temperature before the
onset of neutron triplet pairings, we have switched off the neutron triplet
superfluidity in this plot. The Cas A NS core temperature
$T_{\mathrm{core, A}}$ inferred from the observation for the envelope model Ref.~\cite{Potekhin:1997mn}  with $\eta = 5\times 10^{-13}$ is shown in the gray line, while its
uncertainty is estimated by varying $\eta = 10^{-(8-18)}$
(gray band) \cite{Ho:2014pta}. We find that the predicted core 
temperature falls below $T_{\mathrm{core, A}} \simeq 5 \times 10^8$~K
for $f_a = (\text{a few}) \times 10^{8}$~GeV, and thus $f_a$
smaller than this value is disfavored. We also note that the bound derived in this
manner has a large uncertainty due to the ignorance of the
envelope parameter $\eta$.

\begin{figure}
{\includegraphics[clip, width = 0.45 \textwidth]{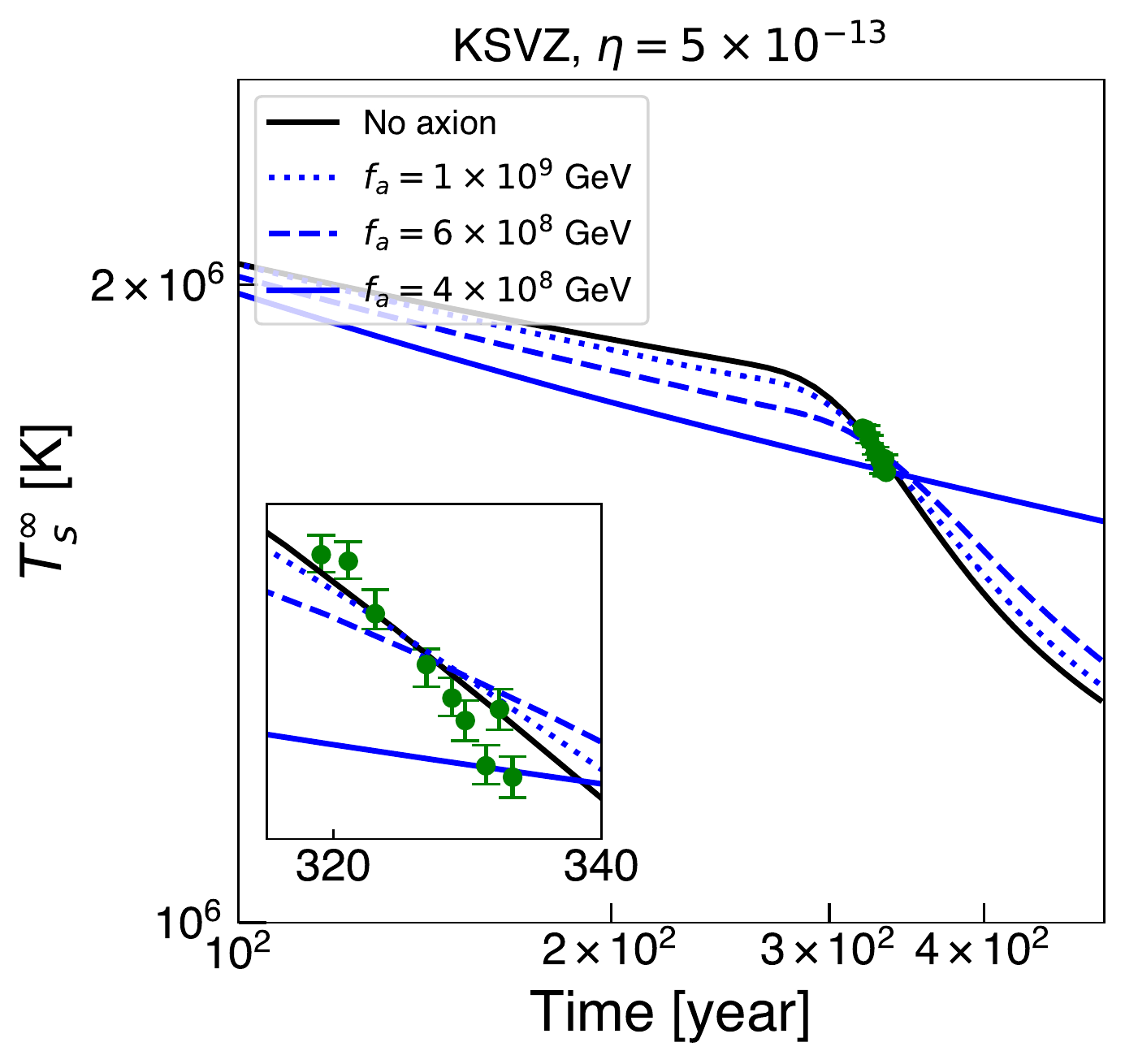}} 
\caption{
Cooling curves compared to observed data.
}
\label{fig:cooling}
\end{figure}

Figure~\ref{fig:cooling} shows the best-fit curves of the red-shifted surface
temperature $T_s^\infty$ for several values of $f_a$ in the KSVZ model
(blue lines), as well as that obtained in the minimal cooling scenario
(black line). For each curve, we vary the neutron triplet
gap parameters and the Cas A NS age to fit the observed data shown
in the green points \cite{Ho:2014pta}, where the envelope parameter is
fixed to be $\eta = 5\times 10^{-13}$ as in \cite{Page:2010aw}. We find that as $f_a$
gets smaller, the NS temperature at $t \simeq 320$~years gets
lower, which then requires a smaller value of $T^{(n)}_c$ and results
in a shallower slope. As a result, the fit gets considerably worse for a
smaller $f_a$. 

\begin{figure}
{\includegraphics[clip, width = 0.45 \textwidth]{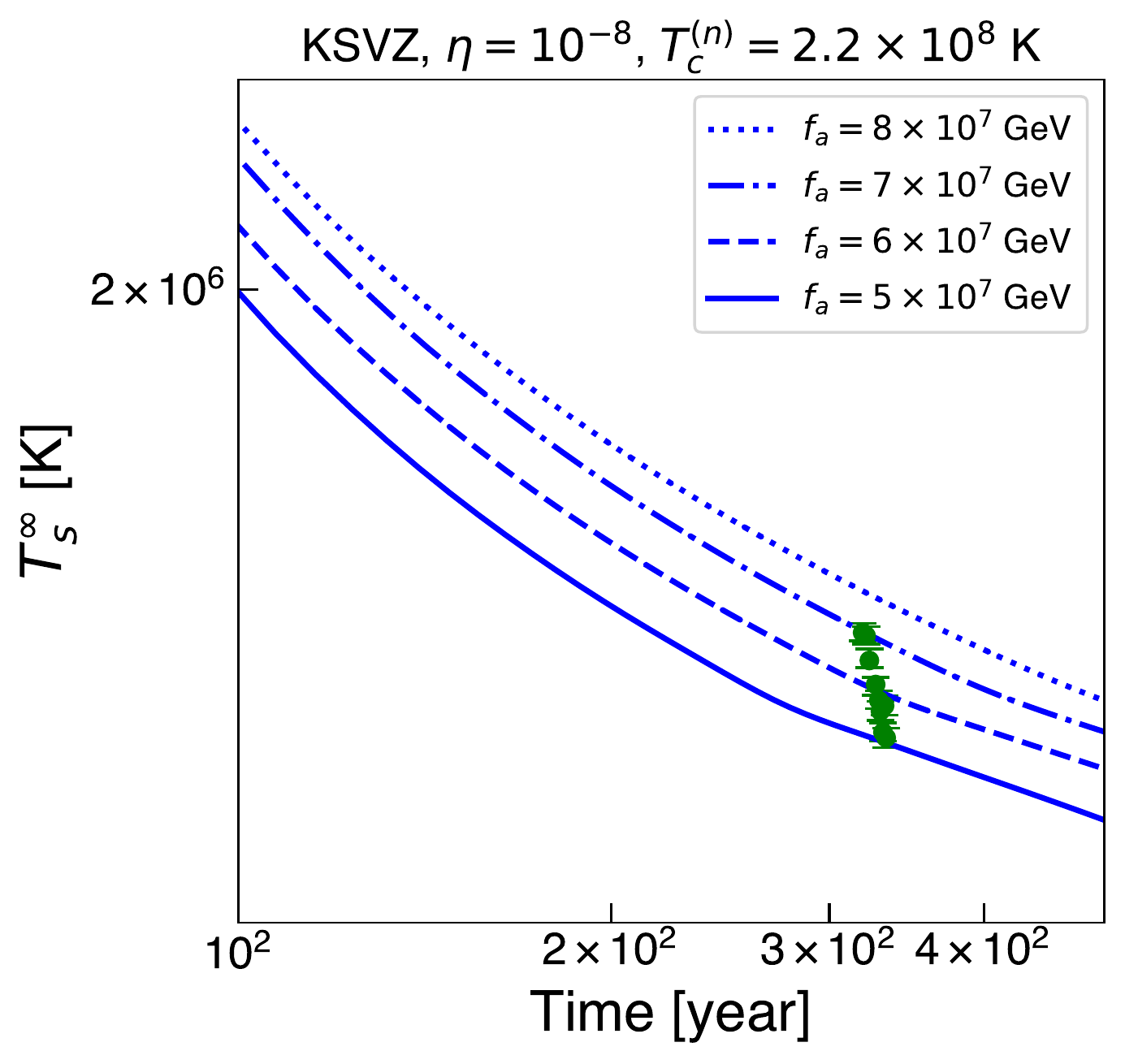}} 
\caption{
Cooling curves of the KSVZ model for $\eta=10^{-8}$.
}
\label{fig:cooling_hieta}
\end{figure}

Finally, we show the lower bound on $f_a$ obtained from our attempt to fit the observed data. If the Cas A NS has an iron envelope with a thin carbon layer ($\eta=5\times 10^{-13}$ as in \cite{Page:2010aw}), 
\begin{align}
f_a&\gtrsim 5~(7)\times10^8~\mathrm{GeV}\quad &\text{KSVZ~(DFSZ)}~,
\label{eq:limitonfa}
\end{align}
where we take $\tan\beta=10$ for the DFSZ model.
The bound on the DFSZ model is comparable to the one on the KSVZ model with $C_n\simeq 0$ because of the large luminosity from proton PBF and the proton-neutron bremsstrahlung as shown in Fig.~\ref{fig:Lumiksvz}. For general couplings, the limit can be roughly estimated by
\begin{align}
f_a&\gtrsim \sqrt{0.9~C_p^2+1.4~C_n^2}
\times10^9~\mathrm{GeV}~.
\end{align}
As a comparison, the bound derived from SN1987A is $f_a\gtrsim 4\times 10^8{~\rm GeV}$ \cite{Raffelt:2006cw} for the KSVZ model, comparable to the bounds from the Cas A NS obtained above.
If the envelope is maximally carbon-rich ($\eta=10^{-8}$) instead,
naively,
 the bound will be weakened by an ${\cal O}(1)$ factor as shown in Fig.~\ref{fig:tvsfa}. 
However, as we increase $\eta$ and hence the thermal conductivity of the envelope, 
the same observed effective temperature corresponds to
a lower core temperature in the NS. This reduces drastically the neutrino luminosity from
the neutron PBF that scales as $T_{\mathrm{core}}^7$ for $T_{\mathrm{core}} \simeq T_c^{(n)}$, making it harder to fit the rapid cooling slope alone. An axion emission may help to cool the NS, but in the KSVZ model the neutrino emission dominates over axion in the neutron PBF process and hence it can be incompatible with the observed rapid cooling. In Fig.~\ref{fig:cooling_hieta}, we plot the cooling curves
of the KSVZ model for a NS with $\eta=10^{-8}$. Due to the large
$\eta$, neutrino emission cannot cool the NS 
to its observed temperature and a sizable axion emission from proton 
with $f_a\lesssim 8\times10^7$~GeV is needed. 
The neutron triplet pairing temperature is set to 
$T_c^{(n)}=2.2\times10^8{~\rm K}$ so that the phase transition occurs 
shortly before the observation. However, we cannot see any rapid 
cooling drop in the curves because the neutrino PBF luminosity 
is suppressed as described above. For a moderate $\eta$ ($\sim 10^{-10}$),
on the other hand, we find that the slope of the cooling data constrains 
$f_a\gtrsim 5\times 10^8~{\rm GeV}$, which is similar to that for 
$\eta = 5\times 10^{-13}$ given in Eq.~\eqref{eq:limitonfa}. 
Thus the limit on the KSVZ model is rather stringent even if 
we allow $\eta$ to vary. For the DFSZ model, $|C_n|$ is non-vanishing 
in general so the axion emission during the neutron triplet-pairing 
phase transition can rapidly cool the Cas A NS via the PBF process 
even when $\eta = 10^{-8}$. According to Fig.~\ref{fig:tvsfa}, 
$f_a\sim 10^8$~GeV is needed to reproduce the observed rapid cooling 
in this case.
We note in passing that such a stellar cooling source may be 
favored by several astrophysical observations 
\cite{Giannotti:2015kwo, *Giannotti:2017hny}, for which our 
new limits 
(or a favored value of $f_a$ in the case of the DFSZ model 
for a large $\eta$) may have important implications. 

To conclude this section, we point out that if $f_a$ is too 
small the axion may have a short mean free path in the NS and 
thus avoid all the limits set above. For the purpose of qualitative 
estimation, we only consider the partial axion decay rate by 
the inverse proton PBF $a\rightarrow\tilde{p}^+ + \tilde{p}^-$ 
that is important to both the KSVZ and DFSZ models. Here,
$ \tilde{p}^\pm$ is the quasi-particle excitation inside a medium 
of proton Cooper pairs. A more careful evaluation is beyond the scope of
this paper. The matrix element
of the related process is given in \cite{Keller:2012yr}, which leads to
\begin{equation}
\Gamma_{a\rightarrow\tilde{p}^+\tilde{p}^-}
\sim 
\frac{m_p^* p_{F} v^2_{F} T}{3\pi f_a^2}\left(\frac{C_p}{2}\right)^2~,
\end{equation}
where $m_p^*$, $p_{F}$, $v_{F}$ are the effective mass, the Fermi momentum, 
and the Fermi velocity of proton in the NS, respectively. 
For $p_F\sim100$~MeV, $m_p^*\sim1$~GeV, $T\sim \Delta_p\sim 1~$MeV,
we need 
\begin{equation}
f_a\gtrsim \left(\frac{C_p}{2}\right)\times 10^6{\rm~GeV}~,
\end{equation}
for the mean free path of axion $l_a=1/\Gamma_{a\rightarrow\tilde{p}^+\tilde{p}^-}$ to be larger than 
$\sim 10~{\rm km}$, the
radius of the Cas A NS.

\section{Conclusion and Discussion}

We have discussed the implications of the Cas A NS rapid cooling for
axion models. It is found that the requirement of
fitting the slope of the Cas A NS cooling curve in accordance with the
temperature evolution from its birth results in a limit of $f_a \gtrsim 5\times 10^8$~GeV if the envelope only has a thin layer of carbon. 

This limit is
stronger than those obtained in the previous studies. For instance, Ref.~\cite{Sedrakian:2015krq} sets 
$f_a \gtrsim (5-10)\times 10^7$ for the KSVZ model without taking the cooling rate of the Cas A NS into account. In Ref.~\cite{Leinson:2014ioa}, it
is argued that the rapid cooling of the
Cas A NS can be explained with $|C_n|/f_a \simeq
\sqrt{0.16}/(10^9~\mathrm{GeV})$ in the KSVZ model---this corresponds to
$f_a \simeq 5\times 10^7$~GeV for $C_n = -0.02$, which is actually in
tension with the observation as shown in Fig.~\ref{fig:tvsfa}. 

Finally, some remarks on the uncertainties of our analysis are in order. First of all, lower cooling rates of the Cas A NS have been reported and the actual cooling rate is still in dispute~\cite{Elshamouty:2013nfa,*Posselt:2013xva}.
In the worst case scenario where the NS is found to cool slowly by future observations, our strong limit on the KSVZ model will no longer hold. However, the conservative limits obtained directly from Fig.~\ref{fig:tvsfa} by assuming maximal $\eta$ are hardly affected since they do not rely on the cooling rate and depend dominantly on the emission from proton. Other mechanisms~\cite{Yang:2011yg, *Negreiros:2011ak, *Blaschke:2011gc, *Noda:2011ag, *Sedrakian:2013xgk, *Blaschke:2013vma, *Bonanno:2013oua, *Leinson:2014cja, *Taranto:2015ubs, *Noda:2015pvn, *Grigorian:2016leu} such as an extended thermal relaxation time have also been proposed to explain the rapid cooling rate and more study on neutron star physics is needed to test them against the minimal cooling scenario that we base our work on. 

Apart from the observation and theoretical uncertainties stated above, the limit on $f_a$
obtained in this work also suffers from the
ignorance of the envelope parameter $\eta$.
{While a maximal $\eta$ parameter is inconsistent with the KSVZ axion model, it is compatible with the DFSZ model with $f_a \sim 10^8$~GeV.}
 Further cooling data of the Cas A NS,
as well as additional observations of direct cooling curves of other
NSs, are of great importance {to verify the rapid cooling of the Cas A NS and} to test the present scenario {against potential alternative
explanations of the rapid cooling of the Cas A NS}, and may allow
us to obtain a more robust limit on the axion.
The analysis performed in
this paper can also be applied to other cooling sources, \textit{e.g.}, heavier axion-like particles and dark photons. {Such an analysis will be given in a future work \cite{HNYZ}. }

\begin{acknowledgments}
\section*{Acknowledgements}
This work was supported in part by the Grant-in-Aid for Scientific
Research A (No.16H02189 [KH]), Young Scientists B
(No.17K14270 [NN]), and Innovative Areas (No.26104001 [KH],
26104009 [KH, JZ]).
The work of KY was supported by JSPS KAKENHI Grant Number JP18J10202. 
\end{acknowledgments}


\bibliography{ref}

\end{document}